# Machine-Learning-Enabled Full-State Reconstruction of Fusion Plasmas from Minimal Sensor Measurements

Maryam Reza*[1], Farbod Faraji**

* Independent researcher

**Department of Computing, Imperial College London, London SW7 2RH, United Kingdom

**Abstract**:

Plasma in nuclear fusion reactors is only partially observable: diagnostics are constrained by limited access, cost, and the harsh plasma environment, while high-fidelity simulations remain prohibitively costly at reactor-relevant scales to address the observability gap. This paper presents an ML model for reconstructing full-domain plasma states from a small number of accessible measurements. The model combines temporal encoding of sparse sensor histories with spatial decoding into complete plasma-field maps. Demonstrations using high-fidelity kinetic simulation data show that multiple coupled plasma quantities can be reconstructed from only a few density sensors, with robustness to sparse and randomly located probes. The approach provides a route toward diagnostic augmentation, real-time state estimation, and data-driven digital twins for fusion-relevant plasmas.

## 1. Introduction:

A central difficulty in fusion plasma monitoring is that the quantities most relevant to performance, confinement, stability, and wall interactions are not fully observable, and often lack direct, real-time measurement. Diagnostics are constrained by limited lines of sight, restricted access to high-load regions, finite spatial and temporal resolution, plasma perturbation by intrusive probes, and the cost and complexity of installing dense diagnostic arrays. As a result, the measurements available during operation represent only a small projection of the full plasma state; thus, critical internal states and the heat-load patterns to the walls and through the exhaust remain effectively "unseen". This motivates the development of novel approaches capable of leveraging sparse, accessible measurements to infer the full-domain plasma state maps information required for interpretation and control.

Physics-based modelling and classical state estimation partly address this problem, but they introduce limitations for real-time use. High-fidelity plasma simulations remain computationally demanding because they must resolve multiscale, nonlinear, and often kinetic dynamics. Conventional compressed-sensing [1] and Bayesian data-assimilation/state-estimation [2,3] approaches also face significant hurdles: (a) the requirement for more probes than is practically feasible; (b) the need for near-optimal sensor placement, remaining highly sensitive to sensor number/location, which is difficult to satisfy in reactor-relevant environments where sensors may be sparse, indirect, non-optimally located, or infeasible to be placed in the most informative regions; and (c) the necessity for accurate, often slow, underlying physics models.

In this contribution, we demonstrate a machine-learning (ML) methodology that reconstructs and forecasts full high-dimensional plasma state maps from *minimal* sensor measurements, overcoming the limitations of the classical techniques. This capability enables access to information beyond what is directly measurable in data-constrained fusion environments using a temporal encoder coupled to a spatial decoder that robustly learns nonlinear spatiotemporal dynamics.

## 2. The machine-learning model architecture

The reconstruction problem is formulated as a mapping from a finite sequence of sparse measurements to a full spatial state [4]. Let the plasma state at a given time contain the spatial distributions of several coupled variables.

[1] **Corresponding author**: m.reza20@alumni.imperial.ac.uk

Only a small number of local measurements are assumed to be available, and these measurements may sample only one physical quantity. The input to the model is a lagged sequence of these sensor values over a finite window K, and the output is the full spatial distribution of all plasma properties at the present time.

The architecture has two coupled components (Figure 1). The temporal encoder uses recurrent neural network - Long Short-Term Memory (LSTM) network in this work – to learn the dynamical information contained in the sensor temporal trajectories. This is important because a single instantaneous sensor value is generally insufficient to determine the full state; a short history contains phase, oscillation, growth, saturation, and transport information. The spatial decoder then maps the learned latent representation to the high-dimensional spatial fields. In this way, the model separates the task into learning temporal dynamics from sparse signals and learning the spatial structure associated with those dynamics.

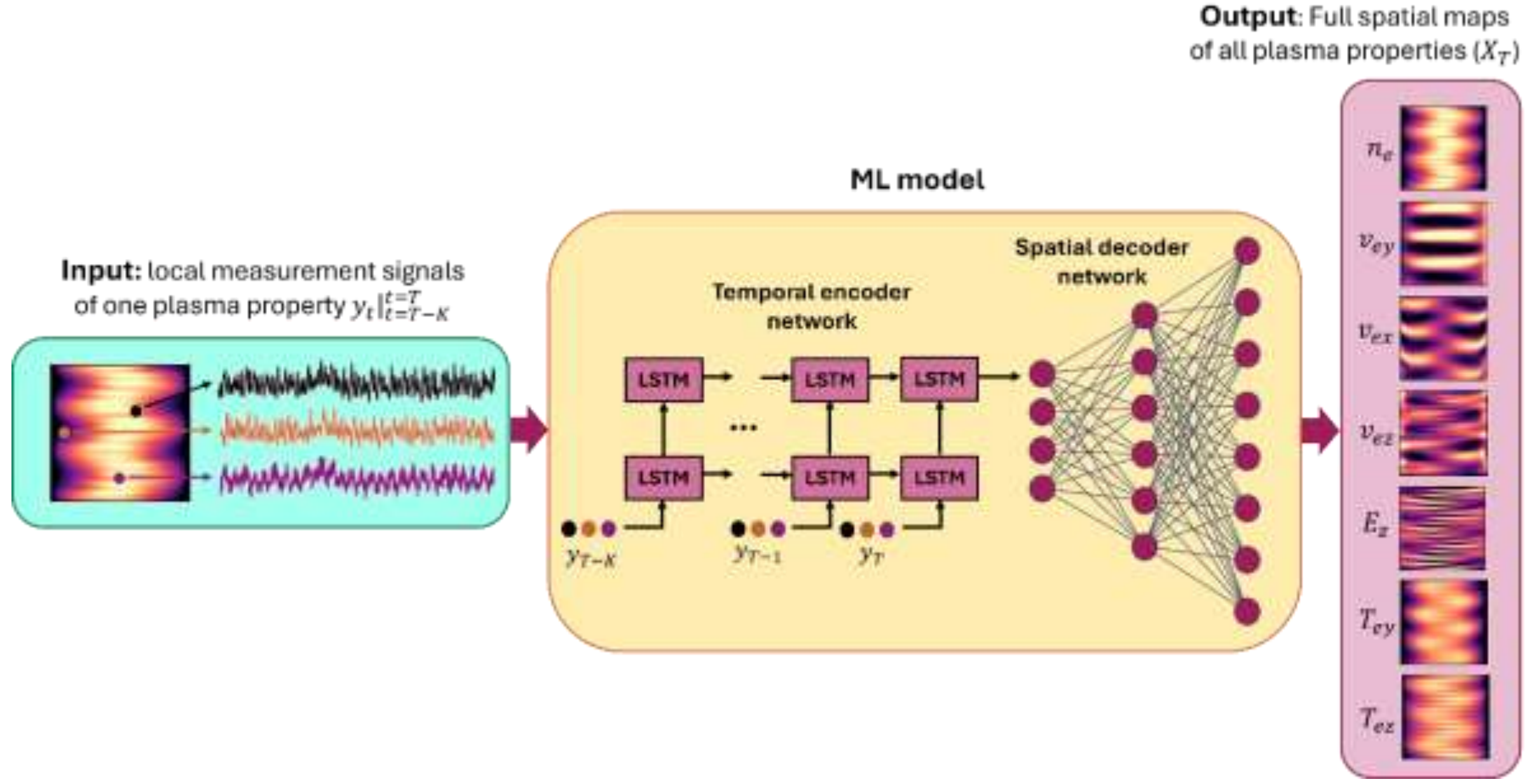

*Figure 1: The architecture for modelling plasma full-state reconstruction from minimal sensor measurements.*

This design has a close physical interpretation. In coupled dynamical systems, time-resolved measurements at a few locations can contain information about the broader spatial state, because the observed signals arise from the same underlying modes, transport processes, and instabilities that shape the full field. Instead of relying on dense spatial measurements at a single instant, it uses time-resolved sensor signals to infer the broader plasma state. In this sense, the availability of temporal information from a few accessible diagnostics is exchanged for spatial information that would otherwise require many distributed sensors and may be impractical to obtain experimentally.

## 3. Demonstration case and results

The model is demonstrated on a benchmark E x B plasma configuration [5], where the plasma is subject to perpendicular electric and magnetic fields. Although the example is not a fusion reactor, it is a representative testbed for fusion-relevant reconstruction complexities as it contains strongly nonlinear plasma dynamics, multiscale instabilities, and complex spatiotemporal coupling between several plasma properties. These are precisely the features that make full-state inference challenging in fusion plasmas.

The ML model is trained on high-fidelity two-dimensional particle-in-cell simulation data on a 256 x 256 grid. The input consists of time histories of length K = 50 from only three randomly placed plasma-density sensors. For demonstration, these sensor signals are extracted from the simulation data as local virtual-probe measurements; in experimental application, they would be replaced by time-resolved signals from real plasma diagnostics or probes. The output consists of full spatial maps of the relevant plasma properties reconstructed simultaneously, including electron number density ($n_e$), electron velocity components ($v_{ex}$, $v_{ey}$, $v_{ez}$), electric field ($E_z$), and electron temperature related to electron random velocity components ($T_{ey}$ and $T_{ez}$). A single measured property – electron number density in this case – is used to infer the remaining unmeasured but dynamically linked quantities.

The reconstructed sample snapshots in Figure 2 show that the ML model accurately captures the full plasma state across all plasma variables from only three input density sensors.

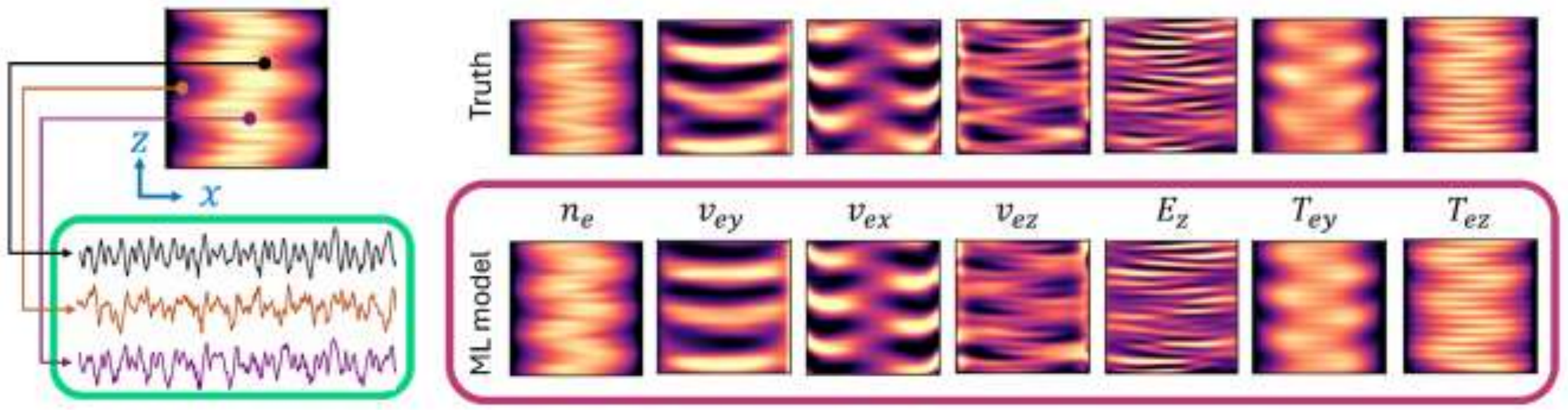

*Figure 2: Sample snapshots of reconstructed spatial distributions of key plasma properties from the trained ML model (lower row), compared against the corresponding simulation ground truth (upper row)* [4].

## 4. Robustness to sensor sparsity and placement

A major advantage of this ML technique is its ability to achieve high-fidelity reconstruction with very few sensors. The model remains effective with as few as three measurements, showing that full-state inference can be obtained without dense diagnostic coverage, which is often impractical in plasma systems

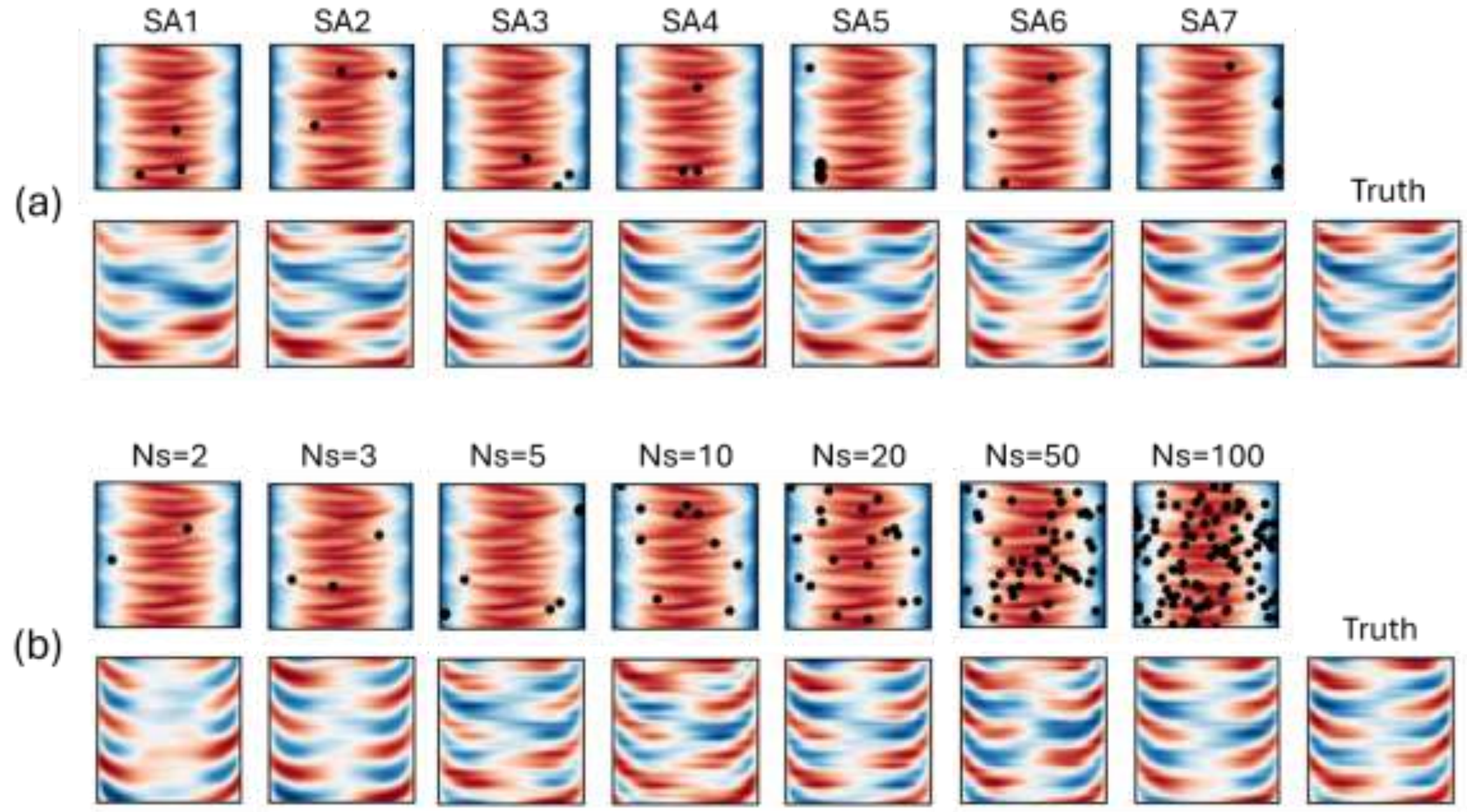

*Figure 3: Effect of (a) sensors arrangement (SA) and (b) sensor numbers ($N_s$) on the ML model reconstruction quality. Black dots indicate input sensor placement* [6].

Another important practical benefit of this technique is its robustness to sensor-placement arrangements over methods whose performance depends on optimal sensor placement. High-accuracy reconstruction is maintained across random probe arrangements, indicating that the model is not strongly dependent on a carefully optimized diagnostic layout. In a real fusion device, sensors often have to be placed where they can survive and where access is available; a method with low placement sensitivity is therefore more deployable.

Figure 3 shows that the reconstructed fields remain in close agreement with the ground truth across different sensor arrangements and sensor counts.

### 7. Relevance to fusion plasma applications

For fusion applications, the central value is the ability to infer hidden state information during operation, real-time monitoring and control. Fusion diagnostics often provide only sparse, local, line-integrated, or boundary-accessible signals, while control and protection require richer knowledge of profiles, gradients, heat-load structures, and internal state evolution. Our ML model can therefore augment diagnostic information, converting sparse signals into richer estimates of the plasma state. This capability can support several use cases. It can augment diagnostics by estimating quantities that are difficult or expensive to measure directly. It can provide state information for feedback control, where rapid estimates are required faster than a high-fidelity simulation can be run. This makes the approach relevant to online monitoring, fast scenario assessment, and data-driven digital-twin workflows [7,8]. It can reduce reliance on repeated expensive simulations during operation, while still using simulations during training and validation. It can also serve as a data-driven predictive engine underlying a fusion reactor digital twin, where real-time experimental data are assimilated into a learned representation of the plasma.

### 8. Practical deployment considerations

A practical implementation requires the training data, diagnostic inputs, and reconstructed outputs to be selected jointly. Input signals should be operationally feasible, reliable and dynamically linked to the hidden state variables of interest, while the reconstructed fields should be chosen according to the application, such as monitoring or control. Uncertainty estimation is also essential. The model should indicate when reconstructions are reliable and when diagnostic signals fall outside the training distribution, particularly during transients, regime changes, faults, or diagnostic degradation. This could be supported through ensembles, probabilistic outputs, residual monitoring, or hybrid ML-physics checks. The framework can also support diagnostic design by evaluating reconstruction accuracy for different sensor numbers and arrangements. In this way, it can help determine how much diagnostic information is needed to meet a target accuracy while balancing cost, accessibility, survivability, and intrusiveness.